\documentclass[journal,12pt,onecolumn,draftclsnofoot]{IEEEtran}

\usepackage[letterpaper, left=1in, right=1in, bottom=1in, top=0.75in]{geometry}
\usepackage{graphicx,tabularx,array,amsmath,amsthm,thmtools}
\usepackage{enumitem}
\usepackage{mathtools}
\DeclarePairedDelimiter{\ceil}{\lceil}{\rceil}
\DeclarePairedDelimiter\floor{\lfloor}{\rfloor}
\usepackage{caption,subcaption}
\usepackage{amsfonts}
\usepackage{bm}
\usepackage{bbm}
\usepackage{makecell}
\usepackage{multirow}
 \usepackage{amssymb}
\usepackage{amsthm}
\usepackage{txfonts}
\usepackage[T1]{fontenc}
\usepackage[scr=dutchcal]{mathalfa}
\let\mathscr\mathbscr

\usepackage{cite}
\usepackage{amsmath}
\usepackage{booktabs}
\usepackage{cuted}
\usepackage{algorithm}
\usepackage{algorithmic}

\allowdisplaybreaks 

\declaretheoremstyle[headfont=\bfseries, 
    bodyfont=\normalfont]{normalhead}

\newtheorem{Theorem}{Theorem}

\newtheorem{Lemma}{Lemma}
\newtheorem{Corollary}{Corollary}
\newtheorem{Remark}{Remark}
\newtheorem{Definition}{Definition}

\DeclareMathOperator*{\argmax}{argmax}

\begin{document}
%
\title{Opportunistic Temporal Fair Mode
Selection and User Scheduling for Full-duplex Systems}

\author{Shahram Shahsavari$^\dagger$,  Farhad Shirani$^\dagger$, Mohammad A. (Amir) Khojastepour$^\diamond$, Elza Erkip$^\dagger$\\
 $^\dagger$NYU Tandon School of Engineering, $^\diamond$NEC Laboratories America, Inc. \\
Emails: $^\dagger$\{shahram.shahsavari,fsc265,elza\}@nyu.edu, $^\diamond$ amir@nec-labs.com }


%


\maketitle

\begin{abstract}
In-band full-duplex (FD) communications --- enabled by recent advances in antenna and RF circuit design ---  has emerged as one of the promising techniques to improve data rates in wireless systems. One of the major roadblocks in enabling high data rates in FD systems is the inter-user interference (IUI) due to activating pairs of uplink and downlink users at the same time-frequency resource block. Opportunistic user scheduling has been proposed as a means to manage IUI and fully exploit the multiplexing gains in FD systems. In this paper, scheduling under long-term and short-term temporal fairness for single-cell FD wireless networks is considered. Temporal fair scheduling is of interest in delay-sensitive applications, and leads to predictable latency and power consumption. The feasible region of user temporal demand vectors is derived, and a scheduling strategy maximizing the system utility while satisfying long-term temporal fairness is proposed. Furthermore, a short-term temporal fair scheduling strategy is devised which satisfies user temporal demands over a finite window-length. It is shown that the strategy achieves optimal average system utility as the window-length is increased asymptotically. Subsequently, practical construction algorithms for long-term and short-term temporal fair scheduling are introduced. Simulations are provided to verify the derivations and investigate the multiplexing gains. It is observed that using successive interference cancellation at downlink users improves FD gains significantly in the presence of strong IUI. 
\end{abstract}

\let\thefootnote\relax\footnotetext{This work is supported by NYU WIRELESS Industrial Affiliates and National Science Foundation grants 1547332 and 1527750.}

%
\IEEEpeerreviewmaketitle

\section{Introduction} \label{sec:intro}
The application of full-duplex (FD) radios enables simultaneous uplink (UL) and downlink (DL) communication over a common frequency band which can potentially lead to significant multiplexing gains \cite{goyal2015full}. Self-interference at the FD base station (BS), as well as inter-user interference (IUI) between the UL and DL users are among major roadblocks in achieving these multiplexing gains. Recent advances in antenna and radio frequency circuit design have led to significant progress in efficient self-interference mitigation (SIM) \cite{harish-SI}. Furthermore, opportunistic scheduling along with successive interference cancellation (SIC) decoding methods at the DL receiver can be used to further reduce IUI \cite{goyal2015full}.



In FD systems, the DL user either treats the UL interference (i.e. IUI) as noise or implements SIC methods to mitigate the IUI \cite{ElGamalLec}. 
Consequently, as shown in Figure \ref{fig:example}, the FD system in which the BS is FD and users are half-duplex (HD) may operate in four modes of operation at each resource block: a) HD-UL mode, where only UL transmission occurs, b) HD-DL, where only a DL user is activated, c) FD-IN, where IUI is treated as noise, and d) FD-SIC, where SIC is used to decode and cancel IUI in DL, 
An opportunistic scheduler selects the active users at each resource block as well as their mode of operation so as to maximize the resulting system utility subject to fairness criteria. In the FD-SIC and FD-IN modes, the scheduler activates pairs of UL and DL users such that the impact of IUI is minimized. 


There has been a large body of research on quantifying fairness in user scheduling. Various criteria on the users' quality of service (QoS) have been proposed to model fairness of the scheduling strategies. For HD systems, scheduling under utilitarian \cite{liu-elsevier,zhang2008opportunistic}, proportional \cite{kelly1998rate,viswanath2002opportunistic}, and temporal  \cite{liu-jsac,kulkarni2003opportunistic} fairness criteria have been studied. Temporal fairness is of interest in delay sensitive applications, where a system with predictable latency may be more desirable than a system with highly variable latency, but potentially higher throughput \cite{shahram-allerton}. Short-term temporal fair schedulers guarantee that each user is activated in at least a predefined fraction of time-slots at each finite scheduling window, whereas under long-term temporal fairness, the temporal demands are met over infinitely large window-lengths. Temporal fair scheduling have been investigated in HD wireless local area networks (WLAN) \cite{joshi2008airtime,issariyakul2004throughput} and HD cellular systems \cite{shahram-letter,shahsavari2018joint}. Furthermore, scheduling under long-term temporal fairness in FD systems is considered in \cite{shahram-FD-asilomar}, where a heuristic scheduler is provided by modifying an optimal temporal fair HD scheduler. However, optimal long-term and short-term temporal fair scheduling in FD systems, which is the topic of this paper, have not been investigated before. In our recent works, we have investigated optimal user scheduling under long-term \cite{general-framework-early-access} and short-term \cite{shahsavari2019fundamental} fairness in non-orthogonal multiple access (NOMA) systems. In this paper, we extend our results to FD systems. Several of the results in \cite{general-framework-early-access} and \cite{shahsavari2019fundamental} depend on the underlying NOMA system constraints which are essentially different in the FD system model considered in the paper as articulated in Sections \ref{sec:FR} and \ref{sec:St}. The main contributions of this paper are as follows:

\begin{figure*}
 \centering \includegraphics[width=0.8\linewidth, draft=false]{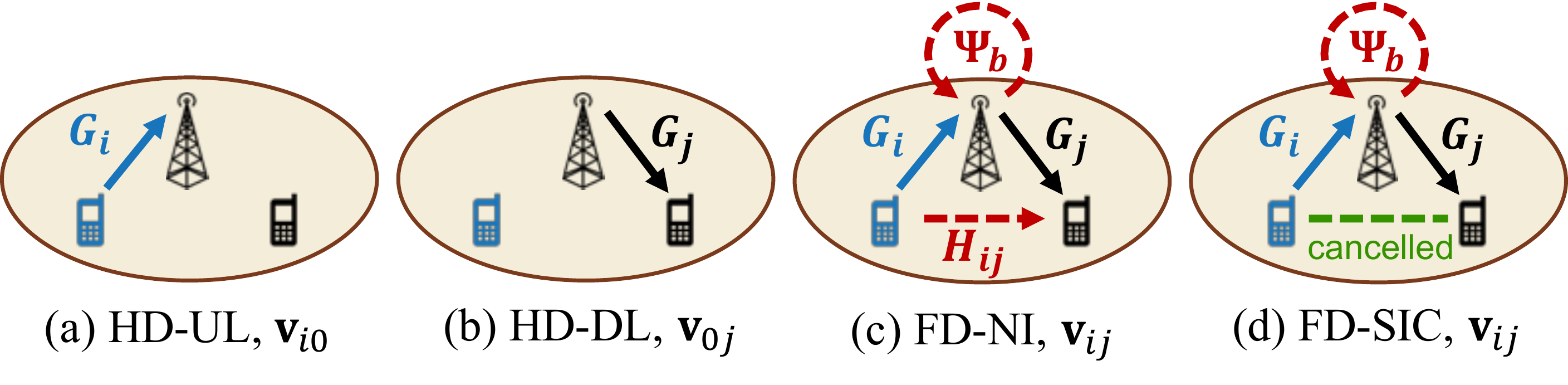}
  \caption{ Figures (a) and (b) show the HD UL and HD DL modes, (c) is the FD mode when IUI is treated as noise, (d) shows the FD mode with SIC in DL, and (e) is the FD mode when the same user is activated in UL and DL.}
 \label{fig:example}
\end{figure*}

\begin{itemize}[leftmargin=*]
         \item We prove that a class of scheduling strategies called threshold based strategies (TBS) achieve optimal system utility under long-term temporal fairness in FD systems. 
         \item We provide an iterative algorithm for construction of optimal TBSs in practice. 
         \item We devise a scheduling strategy under short-term temporal fairness constraints whose average utility is shown to converge to optimal utility as the scheduling window-length grows asymptotically large. 
         \item We present simulation results which verify the derivations under practical FD system models and parameters, and demonstrate the effectiveness of SIC in recovering the FD gains in the presence of strong IUI. 
\end{itemize}

\textit{Notation:}  We represent random variables by capital letters such as $X, U$. Sets are denoted by calligraphic letters such as $\mathcal{X}, \mathcal{U}$.  The set of numbers $\{1,2,\cdots, n\}, n\in \mathbb{N}$ is represented by $[n]$. The vector $(x_1,x_2,\cdots,x_n)$ is represented by $x^n$. Particularly, vectors of length two such as $(x_1,x_2)$ are represented by $\bf{x}$.
The random variable $\mathbbm{1}_{\mathcal{A}}$ is the indicator function of the event $\mathcal{A}$.  Sets of vectors are shown using sans-serif letters such as $\mathsf{V}$. 
 

\section{System Model}
\label{sec:SM}
We consider a single-cell scenario consisting of $n$ HD users and one FD BS. The user set is denoted by $\mathcal{U}=\{u_1,u_2,\cdots,u_n\}$. While this paper considers HD users, the results can be generalized when an arbitrary fraction of the users have FD capability. 
Subsets of distinct users which can be activated simultaneously in UL and DL are called \textit{virtual users}. The set of virtual users is denoted by $\mathsf{V}\subseteq \{\mathbf{v}_{i,j}| i,j\in [n]\cup\{0\}\}$ where $\mathbf{v}_{i,j}$
signifies the instance when the $i$th user is activated in UL and the $j$th user is activated in downlink. If $i$ ($j$) is equal to zero, then the BS operates in the HD UL (DL) mode. 
We note that since the users are HD, $\mathbf{v}_{i,i}\not\in\mathsf{V}, \forall i$. We denote the propagation channel coefficient between user $u_i$ and the BS which captures small-scale and large-scale fading effects at time-slot $t$ by $G_{i,t}$. Similarly, we denote the channel between users $u_i$ and $u_j$ at time-slot $t$ by $H_{i,j,t}, i\not= j$. It is assumed that the channel coefficients $G_{i,t}, i\in [n]$ and $H_{i,j,t}, i,j\in [n]$ are independent over time. Additionally, we assume that the channels are reciprocal.
\subsection{Communication Modes}
As mentioned in Section \ref{sec:intro}, the system may operate in several communication modes. 
\subsubsection{HD-UL and HD-DL modes}
In these modes a single user is scheduled either in UL or DL. Virtual user $\mathbf{v}_{i,0}$ ($\mathbf{v}_{0,j}$) represents the case when user $i$ ($j$) is activated in UL (DL) in HD mode; see Figure \ref{fig:example}(a) (Figure \ref{fig:example}(b)). If $\mathbf{v}_{i,0}$ is activated at time $t$, the SINR for user $u_i$ is
\begin{align}
\text{SINR}^{\text{HD,UL}}_{i,0,t}=\frac{P_{i,t}^{\text{UL}}|G_{i,t}|^2}{N_{\text{UL}}}, 
\end{align}
where $P_{i,t}^{\text{\text{UL}}}$ is the UL transmit power and $N_{\text{\text{UL}}}$ is the UL noise power. Similarly, if $\mathbf{v}_{0,j}$ is activated at time $t$, the SINR for user $u_j$ is as follows,
\begin{align}
\text{SINR}^{\text{HD,DL}}_{0,j,t}=\frac{P_{j,t}^{\text{DL}}|G_{j,t}|^2}{N_{\text{DL}}},
\end{align}
where $P_{j,t}^{\text{DL}}$ is the BS transmit power and $N_{\text{DL}}$ is the DL noise power. 

\subsubsection{FD-IN mode} In this mode, two different users are activated, one in each direction, and the DL user treats IUI as noise. If virtual user $\mathbf{v}_{i,j}, i\not=j$ is activated in FD-IN mode as depicted in Figure  \ref{fig:example}(c), the SINRs of users $u_i$ and $u_j$ are
\begin{align}
&\text{SINR}^{\text{FD-IN,UL}}_{i,j,t}=\frac{P_{i,t}^{\text{\text{UL}}}|G_{i,t}|^2}{P_{j,t}^{\text{DL}}\Psi_b+N_{\text{UL}}},\\
&\text{SINR}^{\text{FD-IN,DL}}_{i,j,t}=\frac{P_{j,t}^{\text{DL}}|G_{j,t}|^2}{P_{i,t}^{\text{UL}}H_{i,j,t}+N_{\text{DL}}}, 
\end{align}
respectively, where $\Psi_b$ is the effective channel between transmit and receive terminals at the BS which is inversely proportional to the level of SIM  at the BS.

\subsubsection{FD-SIC mode}
We assume that an arbitrary fraction of the users are capable of SIC enabling FD-SIC mode. In this mode, two different users are activated, one in each direction, and the DL user performs SIC to cancel IUI.
If virtual user $\mathbf{v}_{i,j}, i\not=j$ is activated in FD-SIC mode (Figure  \ref{fig:example}(d)), the SINR of users $u_i$ and $u_j$ are

\begin{align}
&\text{SINR}^{\text{FD-SIC,UL}}_{i,j,t}=\min\left\{\frac{P_{i,t}^{\text{UL}}|G_{i,t}|^2}{P_{j,t}^{\text{DL}}\Psi_b+N_{\text{UL}}},\frac{P_{i,t}^{\text{UL}}|H_{i,j,t}|^2}{P_{j,t}^{\text{DL}}|G_{j,t}|^2+N_{\text{DL}}}\right\}, \label{eq:fd-sic-ul}
\end{align}
\begin{align}
\text{SINR}^{\text{FD-SIC,DL}}_{i,j,t}=\frac{P_{j,t}^{\text{DL}}|G_{j,t}|^2}{N_{\text{DL}}},
\label{eq:fd-sic-dl}
\end{align}
respectively. FD-SIC imposes two restrictions on the UL transmission rate. First, SIC requires the DL user to be able to decode the UL signal. Second, the BS needs to be able to decode the UL signal. Consequently, the UL rate must be chosen such that the UL signal is decodable at both DL user and the BS. As a result, the SINR of the UL user is defined as the minimum between the SINRs over UL and inter-user channels as in Equation \eqref{eq:fd-sic-ul}. As implied in Equation \eqref{eq:fd-sic-dl}, the DL transmission is interference-free since IUI is cancelled.


\subsection{System Utility}

At any given time-slot, the channel realization, choice of the active virtual user, and the communication mode determine the resulting system utility at that time-slot. In this paper, we take the resulting sum-rate as a measure of the system utility. The following describes the resulting utility from activating each of the virtual users. 

The performance value corresponding to the HD virtual users $\mathbf{v}_{i,0}$  and $\mathbf{v}_{0,j}$ at time-slot $t$ are defined as
\begin{align}
    &R_{i,0,t} =C\left(\text{SINR}^{HD,UL}_{i,0,t}\right),\\
    &R_{0,j,t} =C\left(\text{SINR}^{HD,DL}_{0,j,t}\right),
\end{align}
respectively, where $C(x)=\max\{\log_2(1+x),\gamma_{max}\}$ is the truncated Shannon rate and $\gamma_{max}$ models the maximum feasible spectral efficiency in the system. Virtual user $\mathbf{v}_{i,j} , i\not=j$, can be activated in either of FD-IN and FD-SIC modes if user $u_j$ is capable of SIC. The BS activates the virtual user in the mode which leads to the highest utility. Hence, the utility due to activating $\mathbf{v}_{i,j}, i\not=j$ is given as 
\begin{align}
    R_{i,j,t} =&\max_{X\in\mathcal{X}}\bigg[C\left(\text{SINR}^{X,UL}_{i,j,t}\right) 
    +C\left(\text{SINR}^{X,DL}_{i,j,t}\right)\bigg], 
\end{align}
where, $\mathcal{X}=\{\text{FD-IN,FD-SIC}\}$ if $u_j$ is capable of SIC, and $\mathcal{X}=\{\text{FD-IN}\}$ otherwise.  

In a given time-slot, the system utilities due to activating different virtual users may depend on each other since each user is included in multiple virtual users. However, the performance values in different time-slots are independent of each other due to the independence of channel coefficients over time. The matrix of system utilities due to activating each of the virtual users is called the \textit{performance matrix}. The performance matrix is random and its value depends on the realization of the underlying time-varying channel.
\begin{Definition}[\bf{Performance Matrix}]
The matrix of jointly continuous variables $(R_{i,j,t})_{i,j\in [n]\cup\{0\}}, t\in \mathbb{N}$ is the performance matrix of the virtual users at time $t$. The sequence $(R_{i,j,t})_{i,j\in [n]\cup\{0\}}, t\in \mathbb{N}$ is a sequence of independent matrices distributed identically  according to the joint density $f_{R^{n\times n}}$.
\end{Definition}
\begin{Remark}
In practical scenarios, the performance matrix is a matrix of discrete variables due to discrete modulation and coding schemes. The results of this paper can be extended to the case of discrete performance matrices as in \cite[Sec. VI]{general-framework-early-access}.
\end{Remark}

\subsection{Opportunistic Temporal Fair Scheduling}
 Under temporal fairness, it is required that the fraction of time-slots in which each user is activated in UL or in DL is bounded from below (above). The vector of UL lower (upper) bounds $\underline{w}_{\text{UL}}^n$ ($\overline{w}_{\text{UL}}^n)$ is called the UL lower (upper) temporal demand vector.  Similarly,  the vector  $\underline{w}_{\text{DL}}^n$ ($\overline{w}_{\text{DL}}^n)$ is called the DL lower (upper) temporal demand vector.
The objective is to design a scheduling strategy satisfying the temporal fairness constraints in a given window-length while maximizing the resulting system utility. Accordingly, a scheduling strategy is defined as follows.

\begin{Definition}[\bf{$s$-scheduler}] \label{Def:Strategy}
Consider the scheduling setup parametrized by $(n,\mathsf{V}, \underline{w}_{\text{UL}}^n, \overline{w}_{\text{UL}}^n, \underline{w}_{\text{DL}}^n, \overline{w}_{\text{DL}}^n, f_{R^{n\times n}})$. A scheduling strategy $Q= (Q_t)_{t\in [s]}$  with window-length $s\in\mathbb{N}$ ($s$-scheduler) is a family of (possibly stochastic) functions $Q_t: \mathbb{R}^{n\times n\times t} \to \mathsf{V}, t\in [s]$, where:
\begin{itemize}[leftmargin=*]
    \item { The input to $Q_t, t\in[s]$ is the
sequence of performance matrices $R^{n\times n \times t}$ which consists of $t$ independently and identically distributed matrices with distribution} $f_{R^{n\times n}}$.
\item{ The temporal demand constraints are satisfied:
\begin{align}
&P\left(\underline{w}_{{\text{UL}},i}\leq {A}_{{\text{UL}},i,s}^{Q} \leq \overline{w}_{{\text{UL}},i}, i\in [n]\right)=1,
\\& P\left(\underline{w}_{{\text{DL}},i}\leq {A}_{{\text{DL}},i,s}^{Q} \leq \overline{w}_{{\text{DL}},i}, i\in [n]\right)=1,
\label{Def:tem_fair}
\end{align}}
\end{itemize}
where, the UL and DL temporal shares of user $u_i, i\in [n]$ up to time $t\in [s]$ is defined as
\begin{align}
&A^Q_{{\text{UL}},i,t}=\frac{1}{t}\sum^t_{k=1}\sum_{j=0}^n\mathbbm{1}_{\big\{\mathbf{v}_{i,j}= Q_k(R^{n\times n \times k})\big\}},
\forall i\in [n], t\in [s],\\
&A^Q_{{\text{DL}},j,t}=\frac{1}{t}\sum^t_{k=1}\sum_{i=0}^n\mathbbm{1}_{\big\{\mathbf{v}_{i,j}= Q_k(R^{n\times n \times k})\big\}},
\forall j\in [n], t\in [s].
\label{Def:temp_share}
\end{align}
\end{Definition}

In the context of Definition \ref{Def:Strategy}, an $\infty$-$scheduler$ is a scheduler which satisfies long-term temporal fairness constraints. A scheduling setup where the user temporal shares are required to take a specific value, i.e. ${A}_{i,s}^Q=w_i, i\in [n]$, is called a setup with \textit{equality temporal constraints} and is parametrized by $(n,\mathsf{V}, {w}_{\text{UL}}^n, {w}_{\text{UL}}^n, {w}_{\text{DL}}^n, {w}_{\text{DL}}^n, f_{R^{n\times n}})$.
The following defines the set of feasible window-lengths and temporal demand vectors given a virtual user set.

\begin{Definition}{\bf (Feasible Temporal Demands)}
For a virtual user set $\mathsf{V}$, the window-length $s$,  and temporal demand vector $({w}_{\text{UL}}^n, {w}_{\text{DL}}^n)$ are called feasible if a scheduling strategy satisfying the equality temporal demand constraints exists. The set of all feasible window-lengths and demand vectors $(s,{w}_{\text{UL}}^n, {w}_{\text{DL}}^n)$ is denoted by $\mathcal{S}(\mathsf{V})$. Particularly, the set of feasible demand vectors for asymptotically large window lengths is defined as follows:
\begin{align*}
   \mathcal{S}_{\infty}(\mathsf{V}) =\lim_{t\to \infty} \Big\{\widehat{w}^{2n}|\exists s:\quad  t\leq s \quad \& \quad   (s,\widehat{w}^{2n})\in \mathcal{S}(\mathsf{V})\Big\},
\end{align*}
where $\widehat{w}^{2n}=({w}_{\text{UL}}^n, {w}_{\text{DL}}^n)$. 
 Furthermore, under inequality temporal constraints, the scheduling setup with upper and lower temporal demand vectors $(\underline{w}_{\text{UL}}^n, \overline{w}_{\text{UL}}^n, \underline{w}_{\text{DL}}^n, \overline{w}_{\text{DL}}^n)$ and window-length $s$ is said to be feasible if:
\begin{align*}
     \exists {w}_{\text{UL}}^n,& {w}_{\text{UL}}^n: (s,{w}_{\text{UL}}^n, {w}_{\text{UL}}^n) \in \mathcal{S}(\mathsf{V}) ,
     \\&
     \underline{w}_{\text{UL}}^n\leq {w}_{\text{UL}}^n\leq \overline{{w}}_{\text{UL}}^n,\quad  \underline{w}_{\text{DL}}^n\leq {w}_{\text{DL}}^n\leq \overline{{w}}_{\text{DL}}^n.
 \end{align*}
\end{Definition}

 The average system utility of an $s$-scheduler is:

\begin{Definition}[\bf{System Utility}]
For an $s$-scheduler $Q$:
\begin{itemize}[wide=0pt]
    \item {The average system utility up to time t, is defined as 
\begin{align}
U^Q_t&=\frac{1}{t}\sum^t_{k=1}\sum_{i\in [n]}\sum_{j\in [n]} R_{i,j,k}\mathbbm{1}_{\big\{\mathbf{v}_{i,j}=Q_k(R^{n\times n\times k})\big\}}.
\label{Def:sys_utility}
\end{align}}
\item{The variable $U^Q_s$ is called the average system utility for the $s$-scheduler.}  An $s$-scheduler $Q_s^*$ is optimal if and only if 
$Q_s^*\in\argmax_{Q\in \mathcal{Q}_s} U^Q_{s}$,
where $\mathcal{Q}_s$ is the set of all $s$-schedulers for the scheduling setup. The optimal utility is denoted by $U^*_s$.
\end{itemize}
\end{Definition}
The objective is to study properties of $U^*_s$ and design scheduling strategies achieving the maximum average system utility under temporal fairness constraints.

\section{Feasibility of Temporal Demands} \label{sec:FR}
In this section, we study the set of feasible temporal demand vectors under long-term and short-term fairness constraints. 
We first consider the feasible demand region under long-term fairness constraints (i.e.   $\mathcal{S}_{\infty}(\mathsf{V})$). 
Then, we investigate feasibility under short-term fairness constraints.
\subsection{Feasibility under Long-term Fairness} \label{subsec:feas-long-term}
The following theorem characterizes the feasible temporal demand region under long-term temporal fairness, i.e. $\mathcal{S}_{\infty}(\mathsf{V})$.

\begin{Theorem} \label{thm:feas-long-term}
For the scheduling setup with virtual user set $\mathsf{V}= \big\{\mathbf{v}_{i,j}| i\neq j, i,j\in \{0\}\cup [n]\big\}$, 
the following holds:
\begin{align*}
    ({w}_{\text{UL}}^n,{w}_{\text{UL}}^n)\in \mathcal{S}_{\infty}(\mathsf{V}) &\iff 
 \begin{cases}
        & \sum_{i\in [n]} {w}_{{\text{UL}},i} \leq 1,\\
        &\sum_{i\in [n]} {w}_{{\text{DL}},i} \leq 1,\\
        &\sum_{i\in [n]} {w}_{{\text{UL}},i}+ {w}_{{\text{DL}},i}\geq 1,\\
        & w_{{\text{UL}},i}+w_{{\text{DL}},i}\leq 1. 
    \end{cases}
\end{align*}
\end{Theorem}
The proof is provided in Appendix \ref{thm-prf:feas-long-term}.

\subsection{Feasibility under Short-term Fairness}
In the next step, we consider feasibility under short-term temporal fairness constraints, where the fairness constraints must be satisfied in window-length $s$. The following theorem provides the feasible region.

\begin{Theorem}
\label{th:2}
For the scheduling setup with virtual user set $\mathsf{V}= \big\{\mathbf{v}_{i,j}| i\neq j,  i,j\in \{0\}\cup [n]\big\}$ and 
fairness window-length $s$, temporal demand vectors $({w}_{\text{UL}}^n,{w}_{\text{UL}}^n)$ are feasible, i.e. $(s,{w}_{\text{UL}}^n,{w}_{\text{UL}}^n)\in \mathcal{S}(\mathsf{V})$, if and only if:
\begin{align}
\exists a_{i,j}, a_{0,j}, a_{0,i}\in [s]:
\begin{cases}
& sw_{{\text{UL}},i}=\sum_{j\in [n]} a_{i,j}+a_{i,0}, i\in [n],\\
& sw_{{\text{DL}},j}=\sum_{i\in [n]} a_{i,j}+a_{0,j}, j\in [n],\\
&\sum_{i,j\in[n]}a_{i,j}+ \sum_{k\in [n]}(a_{k,0}+a_{0,k})=s
\\
        & a_{i,i}= 0, i\in [n]\\
        &a_{i,j}\geq 0, i,j\in [n].
    \end{cases}
    \label{eq:feasib}
\end{align}
\end{Theorem}
Note that in Theorem \ref{th:2}, we must have $w_{{\text{UL}},i}=\frac{k_{{\text{UL}},i}}{s}, i\in [n]$, $w_{{\text{DL}},i}=\frac{k_{{\text{DL}},i}}{s}, i\in [n]$, where $k_{{\text{UL}},i},k_{{\text{DL}},i}\in[s], i\in [n]$. The variable $a_{i,j}$ can be viewed as the temporal share of the user $\mathbf{v}_{i,j}$. The forward proof follows by noting that the first two bounds ensure that the temporal fairness constraints are satisfied, the 
third bound must holds since a virtual user must be activated at each time-slot. The fourth set of constraints follows from the fact that the users do not have the FD capability. 
The converse proof follows by constructing a round robin scheduler which activates 
$\mathbf{v}_{i,j}$ for $a_{i,j}$ of the time. The complete proof is provided in Appendix \ref{app:th:2}. We note that the proof depends on the structure of virtual users. Hence, the arguments provided in \cite{shahsavari2019fundamental} for NOMA systems are not directly applicable.

\section{Threshold Based Strategies} \label{sec:St}
In this section, we show that a class of schedulers called threshold based schedulers achieve optimal utility under long-term temporal fairness constraints. Furthermore, we introduce a scheduler which satisfies short-term fairness constraints whose performance converges to optimal performance as the scheduling window-length is increased asymptotically. 

\subsection{Scheduling under Long-term Fairness Constraints}
The following defines threshold based schedulers. 
\begin{Definition}[\bf{TBS}]
\label{Def:TBS}
For the scheduling setup $(n,\mathsf{V}, \underline{w}_{\text{UL}}^n, \overline{w}_{\text{UL}}^n, \underline{w}_{\text{DL}}^n, \overline{w}_{\text{DL}}^n, f_{R^{n\times n}})$ a threshold based strategy (TBS) is characterized by the pair $(\lambda_{\text{UL}}^n, \lambda_{\text{DL}}^n)\in \mathbb{R}^{2n}$. The strategy $Q_{TBS}(\lambda_{\text{UL}}^n, \lambda_{\text{DL}}^n)=(Q_{TBS,t})_{t\in \mathbb{N}}$ is defined as:
\begin{align}
Q_{TBS,t}\big(R^{n\times n\times t}\big)=\argmax_{\mathbf{v}_{i,j}\in\mathsf{V}} ~M\big(\mathbf{v}_{i,j}\,R_{i,j,t}\big), ~t\in \mathbb{N},
\label{Eq:thresh_str}
\end{align}
where $M\big(\mathbf{v}_{i,j}\,R_{i,j,t}\big)=R_{i,j,t}+\lambda_{{\text{UL}},i}+\lambda_{{\text{DL}},j}$ is the `scheduling measure' corresponding to the virtual user $\mathbf{v}_{i,j}$, and $\lambda_{{\text{UL}},0}=\lambda_{{\text{DL}},0}=0$. 
 The resulting temporal shares are represented as $w_{{\text{UL}},i}=A_{{\text{UL}},i}^{Q_{TBS}}, i\in [n]$ and $w_{{\text{DL}},i}=A_{{\text{DL}},i}^{Q_{TBS}}, i\in [n]$. The utility of the TBS is written as $U_{w^n}(\lambda_{\text{UL}}^n, \lambda_{\text{DL}}^n)$. 
The set of threshold based strategies is denoted by $\mathcal{Q}_{TBS}$.
\label{def:U_TBS}
\end{Definition}
The following states that the optimal utility under long-term temporal fairness is achieved using TBSs. 

\begin{Theorem} 
For the scheduling setup $(n,\mathsf{V}, \underline{w}_{\text{UL}}^n, \overline{w}_{\text{UL}}^n, \underline{w}_{\text{DL}}^n, \overline{w}_{\text{DL}}^n, f_{R^{n\times n}})$, assume that $(\underline{w}_{\text{UL}}^n, \overline{w}_{\text{UL}}^n, \underline{w}_{\text{DL}}^n, \overline{w}_{\text{DL}}^n)$ is feasible. Then, there exists an optimal threshold based strategy $Q_{TBS}$.
\label{th:neq:normal}
\end{Theorem}
The proof is provided in Appendix \ref{thm-prf:equality}. Note that the arguments provided in \cite{general-framework-early-access} to prove a similar statement for NOMA systems are not directly applicable since the proof depends on the set of virtual users which is different in FD systems.
The following Corollary provides sufficient optimality conditions which will be used for devising a low complexity algorithm to estimate the optimal thresholds in the next sections.

\begin{Corollary}
For the scheduling setup $(n,\mathsf{V}, \underline{w}_{\text{UL}}^n, \overline{w}_{\text{UL}}^n, \underline{w}_{\text{DL}}^n, \overline{w}_{\text{DL}}^n, f_{R^{n\times n}})$, assume that there exist positive thresholds $(\lambda_{\text{UL}}^n,\lambda_{\text{DL}^n})$ satisfying the complimentary slackness conditions:
\begin{align*}
&i\in [n]: \lambda_{{\text{UL}},i}\left(A^{Q_{TBS}}_{{\text{UL}},i}-\underline{w}_{{\text{UL}},i}\right)=0,
\underline{w}_{{\text{UL}},i}\leq
A^{Q_{TBS}}_{{\text{UL}},i}\leq \overline{w}_{{\text{UL}},i},\\
&j\in [n]:\lambda_{{\text{DL}},j}\left(A^{Q_{TBS}}_{{\text{DL}},j}-\underline{w}_{{\text{DL}},j}\right)=0,
\underline{w}_{{\text{DL}},j}\leq
A^{Q_{TBS}}_{{\text{DL}},j}\leq \overline{w}_{{\text{DL}},j},
\end{align*}
where $Q_{TBS}$ is the TBS corresponding to the threshold vector $(\lambda_{\text{UL}}^n, \lambda_{\text{DL}}^n)$. Then, $Q_{TBS}$ is an optimal scheduling strategy. 
\label{cor:CS}
\end{Corollary}
Note that the complementary slackness conditions in Corollary \ref{cor:CS} are written only in terms of the lower temporal demands. Similar sufficient conditions can be derived in terms of the upper temporal share demands.

\subsection{Practical Construction Algorithms}
The optimal thresholds in TBS depend on the statistics of the performance matrix $R^{n\times n}$, which is typically unavailable in practice. In this section, we propose an online algorithm to find the optimal thresholds in an online fashion only by observing the realization of the performance matrix at each time-slot which can be obtained after channel estimation. 
Algorithm \ref{alg:robbins-monro} constructs an optimal TBS using the complementary slackness conditions provided in Corollary \ref{cor:CS}. The algorithm starts with a vector of initial thresholds (e.g. all-zero thresholds). At time-slot $t$, it chooses virtual user $\mathbf{v}_{i^*,j^*,t}$ to be activated based on the threshold vector $(\lambda_{{\text{UL}},t}^n, \lambda_{{\text{DL}},t}^n)$.  It updates the temporal shares and thresholds based on the scheduling decision at the end of the time-slot (line 2-7). The update rule for the thresholds given in lines 6 and 7 are based on a variation of the Robbins-Monro update described in \cite{general-framework-early-access}. The parameter $c$ is the step-size. Lines (8-23) verify that the temporal demand constraints and dual feasibility conditions are satisfied. The computational complexity of the algorithm is proportional to the number of virtual users which is $O(n^2)$.
\begin{algorithm}[h]
\caption{Heuristic Threshold Optimization in TBS }
\textbf{Initialization}: \small $\lambda_{{\text{UL}},i,1}=0$ , 
$\lambda_{{\text{DL}},j,1}=0$, $i,j\in [n]$

  \begin{algorithmic}[1]
    \FOR {$t \in \mathbb{N}$}
    	\STATE $\mathbf{v}_{i^*,j^*,t}=Q_{t}(\lambda^n_{{\text{UL}},t},\lambda^n_{{\text{DL}},t})$
        \STATE $A^Q_{{\text{UL}},i,t+1}=A^Q_{{\text{UL}},i,t+1}+\frac{1}{t+1}\Big(\mathbbm{1}_{\{i=i^*\}}-A^Q_{{\text{UL}},i,t}\Big),  i\in[n]$
        \STATE $A^Q_{{\text{DL}},j,t+1}=A^Q_{{\text{DL}},j,t+1}+\frac{1}{t+1}\Big(\mathbbm{1}_{\{j=j^*\}}-A^Q_{{\text{DL}},j,t}\Big), j\in[n]$
    	\STATE $\lambda_{min}=\min_{i\in [n]} \{\lambda_{{\text{UL}},i,t},\lambda_{{\text{DL}},i,t}\}$
        \STATE $\lambda_{{\text{UL}},i,t+1}=\lambda_{{\text{UL}},i,t}-c\Big(\lambda_{{\text{UL}},i,t}-\lambda_{min}\Big)\Big(\mathbbm{1}_{\{i=i^*\}}-\underline{w}_{{\text{UL}},i}\Big), i\in [n]$
        \STATE $\lambda_{{\text{DL}},j,t+1}=\lambda_{{\text{DL}},j,t}-c\Big(\lambda_{{\text{DL}},j,t}-\lambda_{min}\Big)\Big(\mathbbm{1}_{\{j=j^*\}}-\underline{w}_{{\text{DL}},j}\Big), j\in [n]$
        \FOR{$i=1$ to $n$}
        	\IF{$\lambda_{{\text{UL}},i,t}=\lambda_{min}$ and $A^Q_{{\text{UL}},i,t+1}<\underline{w}_{{\text{UL}},i}$ }
            	\STATE $\lambda_{{\text{UL}},i,t+1}=\lambda_{{\text{UL}},i,t}+c\Big(\underline{w}_{{\text{UL}},i}-A^Q_{{\text{UL}},i,t+1}\Big)$
            \ENDIF
            \IF{$\lambda_{{\text{UL}},i,t}=\lambda_{min}$ and $\lambda_{min}<0$}
            	\STATE $\lambda_{{\text{UL}},i,t+1}=\lambda_{{\text{UL}},i,t+1}+c$
            \ENDIF
        \ENDFOR
        \FOR{$j=1$ to $n$}
        	\IF{$\lambda_{{\text{DL}},j,t}=\lambda_{min}$ and $A^Q_{{\text{DL}},j,t+1}<\underline{w}_{{\text{DL}},j}$ }
            	\STATE $\lambda_{{\text{DL}},j,t+1}=\lambda_{{\text{DL}},j,t}+c\Big(\underline{w}_{{\text{DL}},j}-A^Q_{{\text{DL}},j,t+1}\Big)$
            \ENDIF
            \IF{$\lambda_{{\text{DL}},j,t}=\lambda_{min}$ and $\lambda_{min}<0$}
            	\STATE $\lambda_{{\text{DL}},j,t+1}=\lambda_{{\text{DL}},j,t+1}+c$
            \ENDIF
        \ENDFOR
    \ENDFOR
  \end{algorithmic}
  \label{alg:robbins-monro}
  
\end{algorithm}

\subsection{Scheduling under Short-term Fairness Constraints}
In this section, we provide a class of scheduling strategies called \textit{augmented threshold based strategies} (ATBS) for FD systems which satisfy hard short-term temporal fairness constraints. More precisely, the strategy satisfies the temporal fairness constraints in a given window-length $s$ with probability one. It is shown that the average utility due to the proposed scheduler converges to the optimal utility as the length of the scheduling window is taken to be asymptotically large. The scheduling strategy has two phases of operation, i) TBS phase, and ii) compensation phase. In the TBS phase, the strategy operates similar to the optimal TBS designed for long-term fairness constraints. In the compensation phase, the strategy activates virtual users in a way so as to ensure that the temporal fairness criteria are satisfied regardless of the resulting utility. The ATBSs are formally defined below. 

\begin{Definition}[\bf{ATBS}]
\label{Def:ATBS}
For the scheduling setup $(n,\mathsf{V}, \underline{w}_{\text{UL}}^n, \overline{w}_{\text{UL}}^n, \underline{w}_{\text{DL}}^n, \overline{w}_{\text{DL}}^n, f_{R^{n\times n}})$  with window-length $s\in \mathcal{S}$, an ATBS is characterized by the pair of vectors $(\lambda_{\text{UL}}^n, \lambda_{\text{DL}}^n)\in \mathbb{R}^{2n}$. The strategy $Q_{ATBS}(s,\lambda_{\text{UL}}^n, \lambda_{\text{DL}}^n)=(Q_{ATBS,t})_{t\in \mathbb{N}}$ is defined as:
\begin{align}
Q_{ATBS,t}\big(R^{n\times n\times t}\big)=\argmax_{\mathbf{v}_{i,j}\in\mathsf{V}_{t}} ~M\big(\mathbf{v}_{i,j}\,R_{i,j,t}\big), ~t\in \mathbb{N},
\end{align}
where $M\big(\mathbf{v}_{i,j}\,R_{i,j,t}\big)$ is the scheduling measure in Definition \ref{Def:TBS}, and $\mathsf{V}_t$ is the 
`feasible virtual user set' at time $t$ and consists of all virtual users $\mathbf{v}_{i,j}$ satisfying the following conditions:
{
\begin{align}
&s-t\geq  \sum_{i\in [n]} \left(\ceil{s\underline{w}_{{\text{UL}},i}}-(t-1)A^Q_{{\text{UL}},i,t-1}- \!\!\!\!\sum_{j\in [n]\cup{\{0\}}}\!\!\!\!\mathbbm{1}_{\mathcal{A}_{i,j}}\right)^+, 
\label{Eq:1}
\\&
s-t\geq  \sum_{i\in [n]} \left(\ceil{s\underline{w}_{{\text{DL}},i}}-(t-1)A^Q_{{\text{DL}},i,t-1}- \!\!\!\!\sum_{j\in [n]\cup{\{0\}}}\!\!\!\!\mathbbm{1}_{\mathcal{A}_{i,j}}\right)^+,
\label{Eq:2}
\\
&
s-t \leq \sum_{i\in [n]} \left(\floor{s\overline{w}_{{\text{UL}},i}}-(t-1)A^Q_{{\text{UL}},i,t-1}- \sum_{j\in [n]\cup{\{0\}}}\!\!\!\!\mathbbm{1}_{\mathcal{A}_{i,j}}\right) \notag
\\
&\qquad+\sum_{i\in [n]} \left(\floor{s\overline{w}_{{\text{DL}},i}}-(t-1)A^Q_{{\text{DL}},i,t-1}-\!\!\!\! \sum_{j\in [n]\cup{\{0\}}}\!\!\!\!\mathbbm{1}_{\mathcal{A}_{i,j}}\right),
\label{Eq:3}
\\
&s-t\geq \max_{i\in [n]}\Bigg( \Big(\ceil{s\underline{w}_{{\text{UL}},i}}-(t-1)A^Q_{{\text{UL}},i,t-1}- \!\!\!\!
\sum_{j\in [n]\cup{\{0\}}}\!\!\!\!
\mathbbm{1}_{\mathcal{A}_{i,j}}\Big) \notag
\\&\qquad+\Big(\ceil{s\underline{w}_{{\text{DL}},i}}-(t-1)A^Q_{{\text{DL}},i,t-1}- \sum_{j\in [n]\cup{\{0\}}}\!\!\!\!\mathbbm{1}_{\mathcal{A}_{i,j}}\Big)\Bigg),
\label{Eq:4}
\end{align}}
where $x^+= x\times \mathbbm{1}_{x\geq 0}$, $\mathcal{A}_{i,j}=\{\mathbf{v}_{i,j}= Q_{TBS,t}(R^{n\times n\times t})\}$, $Q_{TBS}$ is the TBS with threshold vector $(\lambda_{\text{UL}}^n, \lambda_{\text{DL}}^n)$.
\label{def:U_ATBS}
\end{Definition}

For a given pair of threshold vectors $(\lambda_{\text{UL}}^n, \lambda_{\text{DL}}^n)$ the steps in the corresponding ATBS strategy is described in Algorithm \ref{alg:atbs}. In this algorithm, at each time-slot Equations \eqref{Eq:1} and \eqref{Eq:2} ensure that the uplink and downlink lower temporal demands can be satisfied if the TBS with threshold vector $(\lambda_{\text{UL}}^n, \lambda_{\text{DL}}^n)$ is used in the next time-slot. Equation \eqref{Eq:3} ensures the satisfaction of the upper temporal demands. Equation \eqref{Eq:4} ensures that the users 
are not required to be scheduled in UL and DL simultaneously in the remaining time-slots in order to satisfy the temporal demands. 
\begin{algorithm}[t]
\caption{Augmented Threshold Based Strategy}
\small
\begin{algorithmic}[1]
\STATE $\mathsf{V}_0=\mathsf{V}$
\FOR {$t=1$ to $s$ with step-size $1$ }
    \STATE $\mathsf{V}_t=\phi$
    \FOR {$i,j: \mathbf{v}_{i,j} \in \mathsf{V}_{t-1} $}
        \IF { Equations \eqref{Eq:1}, \eqref{Eq:2}, \eqref{Eq:3} and \eqref{Eq:4} are satisfied}
            \STATE $ \mathsf{V}_{t} = \mathsf{V}_{t} \cup \{\mathcal{V}_j\}$
        \ENDIF
    \ENDFOR
    \STATE $Q_{ATBS,t}\big(R^{n\times n\times t}\big)=\argmax_{\mathbf{v}_{i,j}\in\mathsf{V}_{t}} ~M\big(\mathbf{v}_{i,j},R_{t,i,j}\big)$ 
\ENDFOR
  \end{algorithmic}
  \label{alg:atbs}
\end{algorithm} 
The following Theorem shows that for asymptotically large scheduling window-lengths, the utility of the ATBSs converge to the optimal utility which is achieved by TBSs. 

\begin{Theorem} \label{Th:Conv}
For the scheduling setup $(n,\mathsf{V}, \underline{w}_{\text{UL}}^n, \overline{w}_{\text{UL}}^n, \underline{w}_{\text{DL}}^n, \overline{w}_{\text{DL}}^n, f_{R^{n\times n}})$ let $({\lambda^*}_{\text{UL}}^n, {\lambda_{\text{DL}}^*}^n)$ be the threshold vectors of the TBS which achieves optimal average utility under long-term fairness constraints (i.e. $s\to \infty)$ and $({w_{\text{UL}}^*}^n,{w_{\text{DL}}^*}^n)$ the corresponding temporal share vectors. Then, $\lim_{s\to \infty} U^*_s = U^*$.
\end{Theorem}
The proof follows by similar arguments as in the proof of Theorem 4 in \cite{shahsavari2019fundamental} and is omitted due to space limitations. 

\section{Simulation Results} \label{sec:sim}
We consider a single $50$ m $\times$ $50$ m square cell with a FD BS in the center and four HD users distributed around the BS with an exclusion of central disk with radius $r_{min}=5$ m. This model describes indoor scenarios such as a floor in an office building. 
We adopt channel model of indoor RRH/Hotzone scenario from \cite{3gpp-channel}. We assume that there are no upper temporal demand constraints. To investigate the impact of user distribution, we consider two models: \textit{i}) \textit{uniform model} where the users are distributed uniformly inside the cell and \textit{ii}) \textit{hotspot model} where there are $n_h$ randomly located hotspots within the cell and $n/n_h$ users are distributed uniformly within a circle of radius $10$ m around each hotspot \cite{shahram-FD-asilomar}. Table \ref{tab:sim-param} lists the simulation parameters. The user SINRs are modeled as described in Section \ref{sec:SM} and the network utility is assumed to be the sum-rate. At each time-slot prior to the scheduling, a max-min power optimization is performed for each virtual user including two users and for both FD-IN and FD-SIC modes \cite{zhang2018max}. For a given virtual user and a mode of operation, we find UL and DL transmit powers which maximizes the minimum individual user rates in that virtual user. 
Assuming a non-line of sight (NLOS) channel, maximum UL and DL transmit powers are chosen such that the average SNR of $0$ dB is achievable when a single user is active on the boundary of the cell, i.e. at $d=50\sqrt{2}$ m. 

\begin{table}[h]
\centering
\caption{Simulation parameters}
\scalebox{1}{
\begin{tabular}{ll} \toprule
$\bf Parameter$ 				 & $\bf Value$    \\ \midrule
Bandwidth      					 & $10$ MHz                      \\
Noise spectral density     	     & $-174$ dBm/Hz                  \\ 
Noise figure                     & BS: 8 dB, user: 9 dB\\
Number of hotspots               & $1,2$                      \\ 
Self-interference mitigation   &  $60$ dB, $80$ dB, $100$ dB \\

\makecell[l]{Log-normal shadowing \\ standard seviation}  & LOS: 3 dB, NLOS: 4 dB \\
Path loss in dB ($d$ in km)      	 & \makecell[l]{LOS: $89.5+16.9\log_{10}(d)$\\ NLOS: $147.4+43.3\log_{10}(d)$}   \\ 
Small-scale fading model			 & Rayleigh block fading \\
Maximum spectral efficiency          & $6$ bps/Hz            \\\bottomrule
\end{tabular}}
\label{tab:sim-param}
\vspace{-5pt}
\end{table}

\subsection{Long-term Fairness}
\begin{figure}[h]
 \centering \includegraphics[width=0.6\linewidth, draft=false]{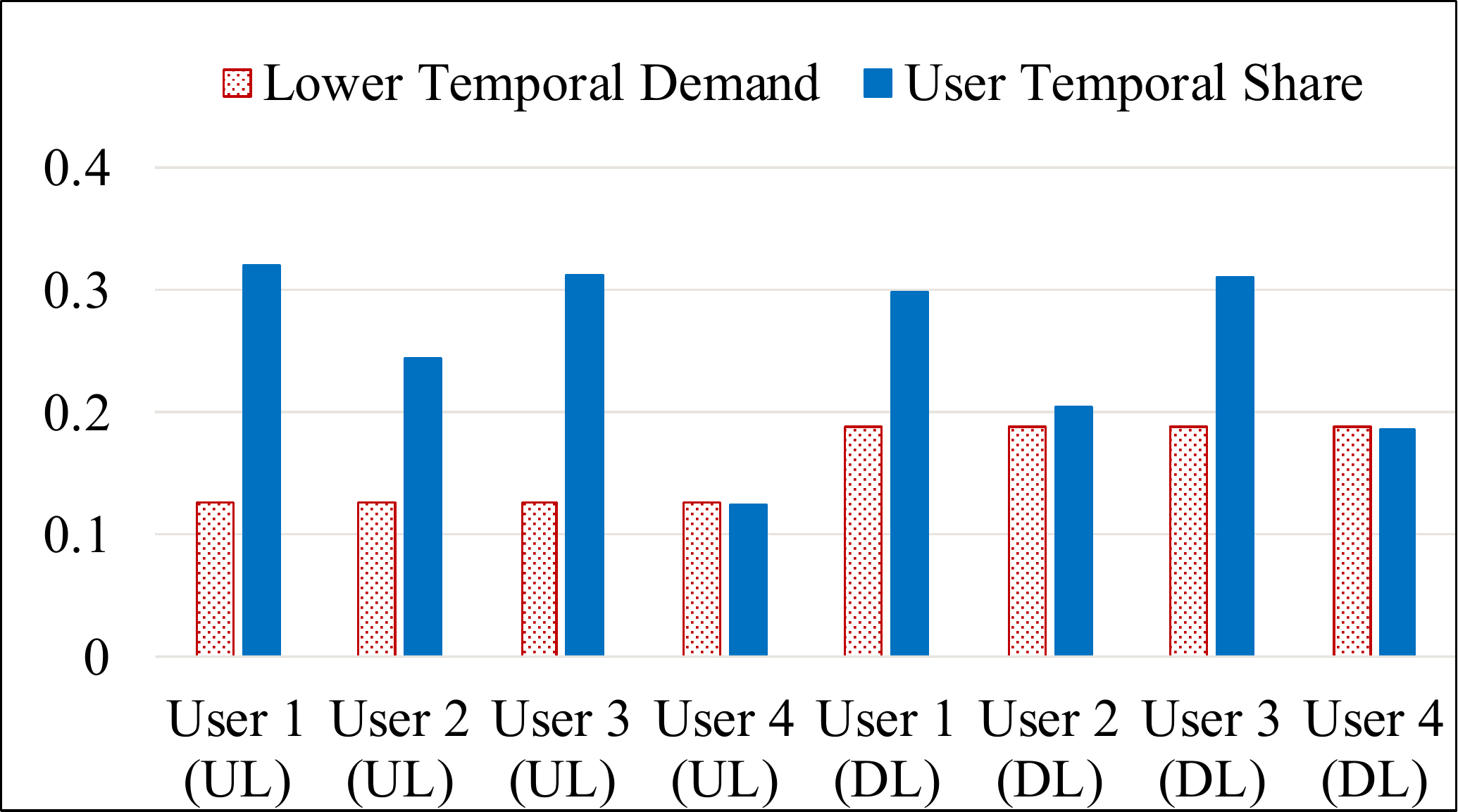}
  \caption{Long-term temporal share of the users versus their lower temporal demands.}
 \label{fig:temp-share}
 \vspace{-5 pt}
\end{figure}

In this section, we consider long-term temporal fairness where $s\to \infty$ and apply Algorithm \ref{alg:robbins-monro} to find the optimal thresholds in TBS. The step-size $c$ is taken to be $0.001$. First, we investigate the capability of Algorithm \ref{alg:robbins-monro} in satisfying long-term temporal fairness constraints. We assume that $\underline w_{{\text{UL}},i}=2/16, \underline w_{{\text{DL}},i}=3/16, i \in[4]$. One can easily check the feasibility of the fairness constraints using Theorem \ref{thm:feas-long-term}. Figure \ref{fig:temp-share} illustrates the temporal shares of the users in UL and DL directions after $4\times 10^6$ time-slots. It can be seen that the temporal demand constraints are satisfied.

Next, we study the impact of availability of FD BS on the system throughput when the user distribution is uniform. We assume that the users are not able to perform SIC, hence the available modes are HD-UL, HD-DL, and FD-IN. We use a HD system (HD BS and users) as a base-line to evaluate the FD gains in the system throughput. 
Furthermore, as a benchmark to FD-IN, we use the heuristic temporal fair scheduler proposed in \cite{shahram-FD-asilomar}. We do not provide the details due to the lack of space and refer the reader to \cite{shahram-FD-asilomar}.
According to \cite{shahram-FD-asilomar}, this heuristic scheduler requires $\sum_i \underline w_{{\text{UL}},i}+\sum_j \underline w_{{\text{DL}},j}\leq 1$, since it uses an underlying HD scheduler. Consequently, there is a set of feasible temporal demands in $\mathcal{S}_{\infty}(\mathsf{V})$ characterized in Section \ref{subsec:feas-long-term}, which are not achievable by this heuristic method whereas according to Theorem \ref{th:neq:normal}, the optimal TBS provided in Section \ref{sec:St} can achieve any choice of temporal demands belonging to $\mathcal{S}_{\infty}(\mathsf{V})$. Furthermore, the heuristic algorithm cannot guarantee upper temporal demands unlike our proposed TBS.

\begin{figure}[h]
 \centering \includegraphics[width=0.5\linewidth, draft=false]{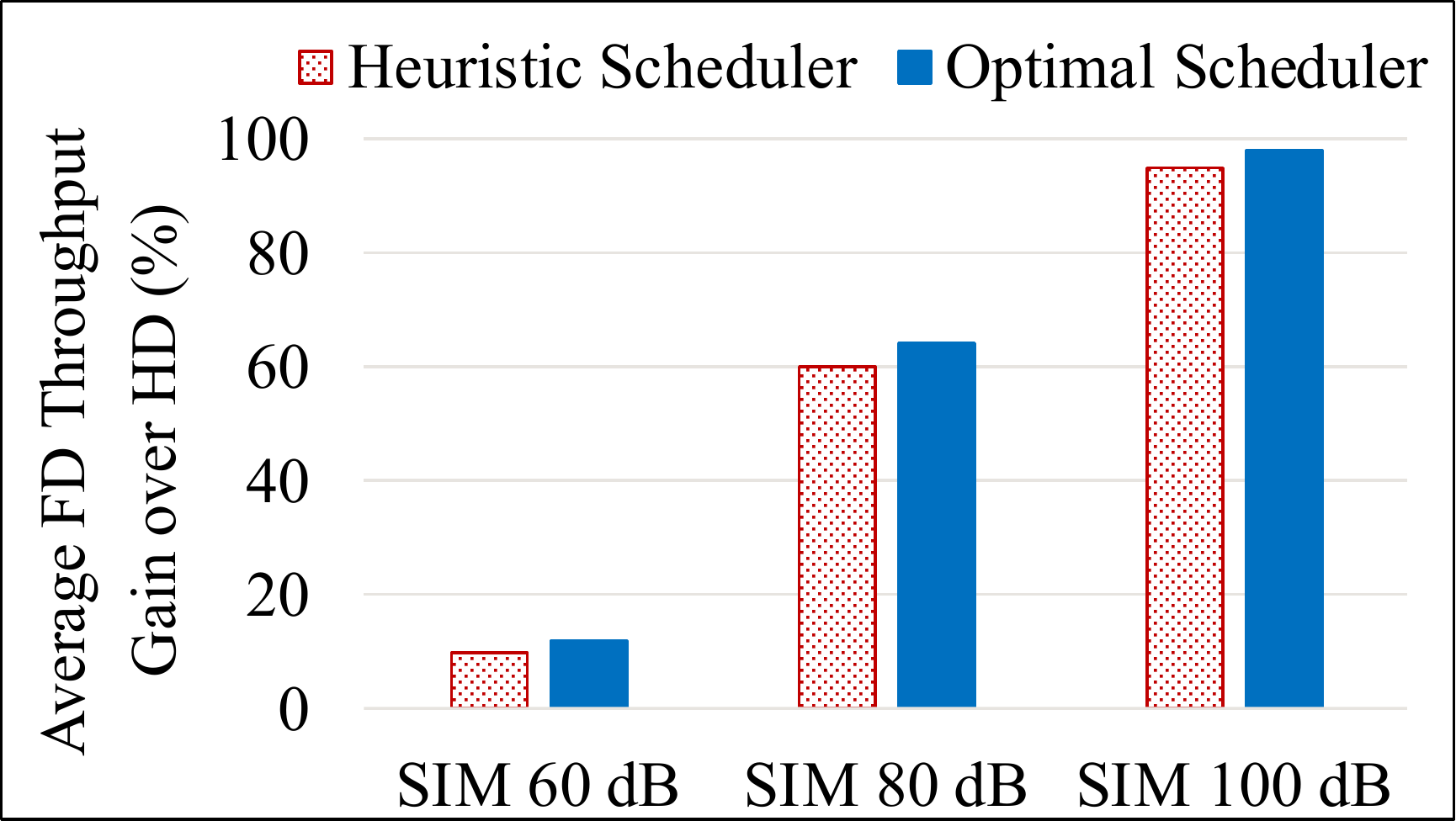}
  \caption{The average gain in the system throughput when comparing various schedulers with the base-line HD scheduler for different levels of SIM at the BS.}
 \label{fig:FD-gain1}
\end{figure}

Figure \ref{fig:FD-gain1} illustrates the average percentage gain in the system throughput when comparing various schedulers with the base-line HD scheduler for different levels of SIM at the BS. We assume that $\underline w_{{\text{UL}},i}=\underline w_{{\text{DL}},i}=1/8, i \in[4]$. Note that these temporal demands are feasible for the base-line HD as well as the heuristic scheduler. We observe that both the optimal and the heuristic scheduler lead to significant improvements for large enough values of SIM at the BS. Additionally, this improvement increases with SIM level at the BS. The reason is that higher SIM leads to less interference for UL reception which in turn improves UL performance value in FD-IN mode.

\begin{figure}[t]
 \centering \includegraphics[width=0.5\linewidth, draft=false]{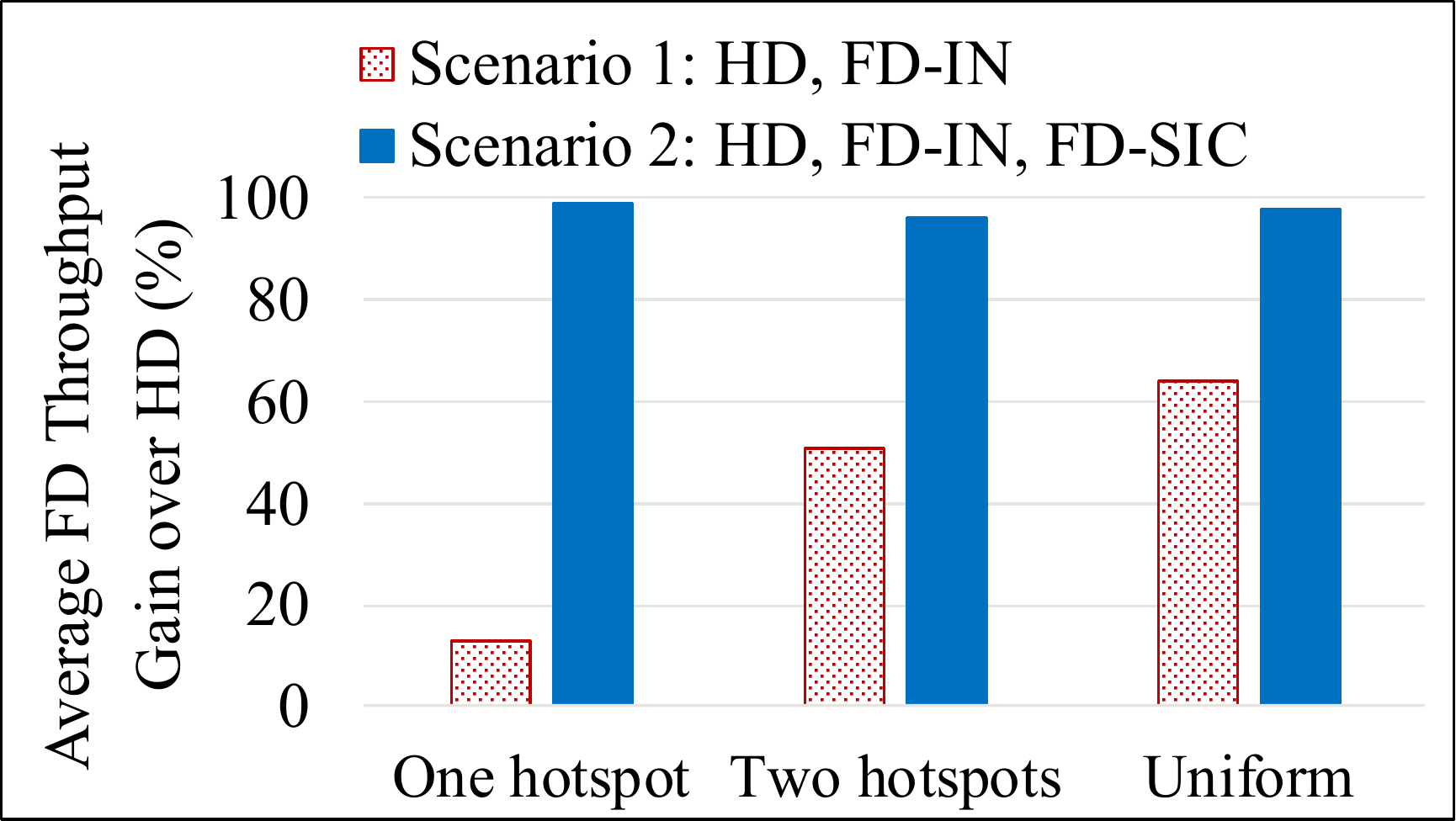}
  \caption{The average system throughput gain when comparing optimal TBS and the base-line HD scheduler for different user distribution models.}
 \label{fig:Fd-gain2}
  \vspace{-5 pt}
\end{figure}

Next, we investigate the impact of user distribution on the multiplexing gains provided by FD operations. We assume that SIM is 80 dB at the BS. Moreover, we consider two scenarios. In the first scenario (Scenario 1), we assume that the users cannot perform SIC. Hence the available modes are HD-UL, HD-DL, and FD-IN. In the second scenario (Scenario 2), we assume that all modes namely HD-UL, HD-DL, FD-IN, and FD-SIC are available. We consider the optimal TBS in both scenarios, where Algorithm \ref{alg:robbins-monro} is used to find the corresponding thresholds. Figure \ref{fig:Fd-gain2} illustrates the average system throughput gain when comparing optimal TBS and the base-line HD scheduler for different user distributions. We observe that when users are located around a single hotspot, the throughput gain achieved in Scenario 1 is limited since IUI is strong and the scheduler tends to use HD-UL and HD-DL modes more frequently. In contrary, the throughput gain is close to $100\%$ in Scenario 2 since the DL user can cancel IUI in the FD-SIC mode which incentivizes the scheduler to select the pairs in this mode more frequently. When there are two hotspots, the throughput gain is higher in Scenario 1 as UL and DL users can be chosen from different hotspots. However, Scenario 2 still leads to higher improvements. An interesting observation from Figure \ref{fig:Fd-gain2} is that SIC is beneficial even if the users are distributed uniformly. The reason is that when SIM level is not very large (e.g. less than 80 dB), performing SIC can improve DL rate at no cost for the UL rate since the UL channel will be the bottleneck for the UL transmission rate rather than the inter-user channel. In other words, with a high probability the first term in \eqref{eq:fd-sic-ul} will be the minimum of the two terms.

\subsection{Short-term Fairness}
\begin{figure}[t]
 \centering \includegraphics[width=0.5\linewidth, draft=false]{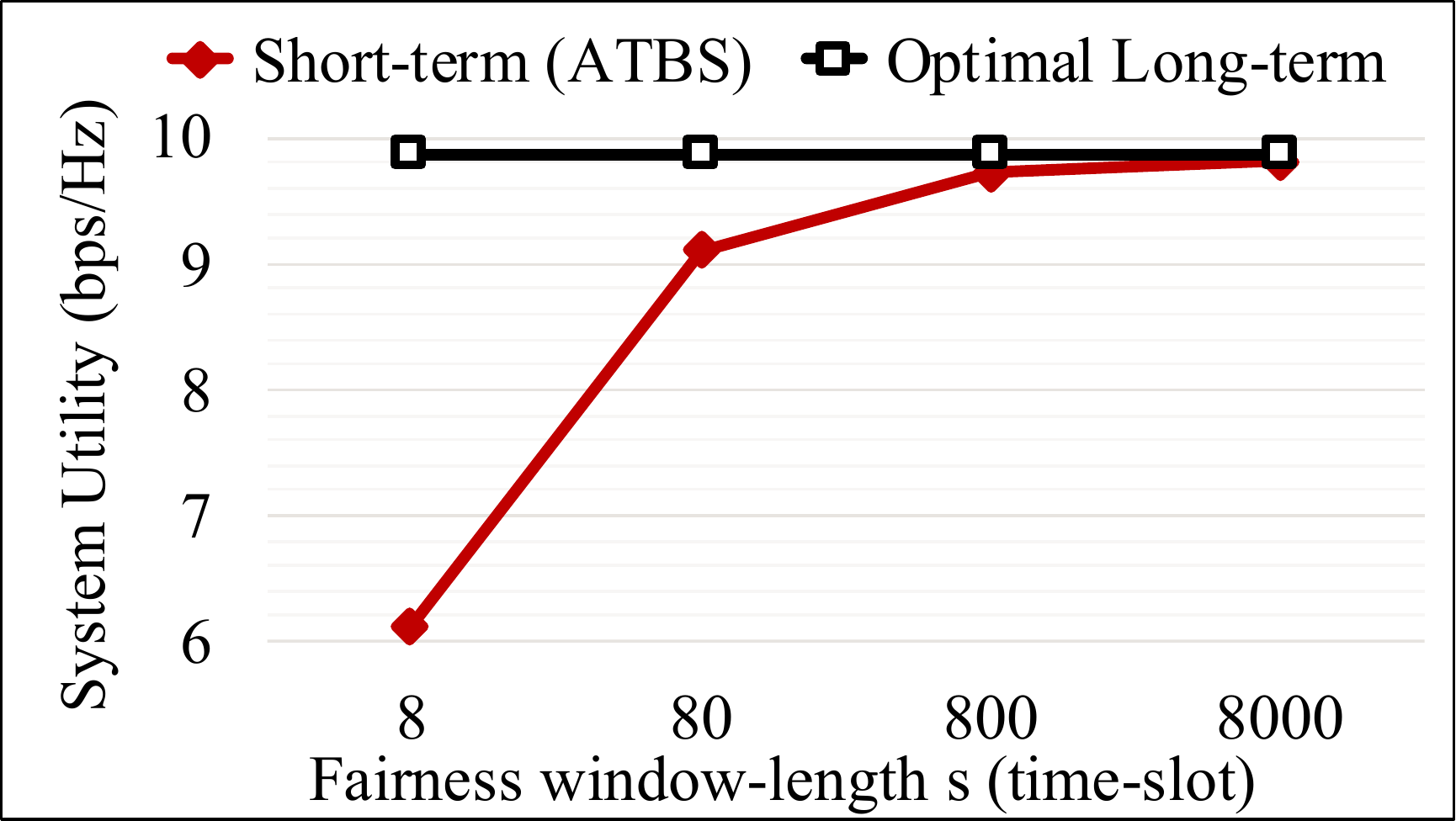}
  \caption{System utility as a function of window-length $s$ for the proposed ATBS using the same thresholds as the optimal TBS.} 
 \label{fig:short-term}
 \vspace{-15 pt}
\end{figure}

Next, we consider a limited fairness window-length $s$ and use Algorithm \ref{alg:atbs} to ensure the short-term fairness. We assume that $\underline w_{{\text{UL}},i}=\underline w_{{\text{DL}},i}=1/8, i \in[4]$. Furthermore, we consider $s\in\{8,80,800,8000\}$. It is straightforward to show that these window-lengths are feasible. Figure \ref{fig:short-term} depicts the average system utility (throughput) as a function of fairness window-length $s$ for ATBS, described in Algorithm \ref{alg:atbs}, using the same thresholds as the optimal TBS. As a benchmark, Figure \ref{fig:short-term} also illustrates the optimal long-term system utility $U_{\infty}^*$ corresponding to the optimal long-term fair scheduling (i.e. optimal TBS). Note that optimal long-term utility is not a function of window-length $s$. It is not difficult to show that $U_{\infty}^*$ provides an upper-bound for the utility of ATBS. We observe that the utility of ATBS approaches the optimal long-term utility as window-length increases, confirming Theorem \ref{Th:Conv}. Furthermore, we can see that the gap between the utility of ATBS and optimal long-term utility is small even for relatively small window-lengths such as $s=80$.

\section{Conclusion}
In this paper, we have studied opportunistic mode selection and user scheduling in single-cell FD systems under short-term and long-term temporal fairness constraints. We have proved that a class of scheduling strategies called threshold based strategies achieve optimal system utility under long-term temporal fairness. Furthermore, we have provided a low-complexity online algorithm for construction of the optimal schedulers under long-term temporal fairness.  Additionally, we have provided a scheduling strategy under short-term temporal fairness constraints whose average utility is shown to converge to optimal utility as the scheduling window-length grows asymptotically large. Simulation results have demonstrated the effectiveness of the proposed scheduling algorithms. A natural extension to this work is multi-cell scheduling in FD systems. The methods proposed here may be extended and applied to centralized and distributed FD systems. Particularly, scheduling for multi-cell FD systems with limited base station cooperation is an interesting avenue for future work. Another avenue is to consider FD capability for the users which can potentially improve the performance in the scenarios where the user distribution is concentrated around a few hotspots and SIC is not available.

\begin{appendices}

\section{Proof of Theorem \ref{thm:feas-long-term}} \label{thm-prf:feas-long-term}
We provide an outline of the proof. 
First we prove the forward result (i.e. any feasible $({w}_{\text{UL}}^n,{w}_{\text{UL}}^n)$ must satisfy the bounds provided in the theorem statement.). The first bound $\sum_{i\in [n]} {w}_{{\text{UL}},i}$ $ \leq 1$ must hold since at most one user can be activated in UL at each resource block. The second bound $\sum_{i\in [n]} {w}_{{\text{DL}},i} \leq 1$ must hold since at most one user can be activated in DL at each resource block. The third bound $\sum_{i\in [n]} {w}_{{\text{UL}},i}+\sum_{i\in [n]} {w}_{{\text{DL}},i}\geq 1$ must be satisfied since at least one user is activated at each resource block either in UL or DL. The Fourth bound $w_{{\text{UL}},i}+w_{{\text{DL}},i}\leq 1$ must hold for all HD users since they cannot be activated in both UL and DL at any resource block. Next, we prove the converse result (i.e. any $({w}_{\text{UL}}^n,{w}_{\text{UL}}^n)$ satisfying the bounds provided in the theorem statement is feasible.). As a first step, we prove the theorem for $k=0$, when all users are HD. We provide a detailed description for the two-user case. For more than two users, we provide an outline of the proof. Assume that $n=2$, it is straightforward to see that a round-robin scheduler satisfying the temporal demands exists. To elaborate, let $\alpha= \sum_{i\in [n]} {w}_{{\text{UL}},i}+\sum_{i\in [n]} {w}_{{\text{DL}},i}- 1$. The variable $\alpha$ indicates the fraction of resource blocks the BS operates in FD mode. The round robin scheduler activates $\mathbf{v}_{1,2}$ for the first $a_1=\min(w_{{\text{UL}},1},w_{{\text{DL}},2},\alpha)$ fraction of the resource blocks. Next, it activates $\mathbf{v}_{2,1}$ for $a_2=\min(w_{{\text{UL}},2},w_{{\text{DL}},1}, \alpha-a_1)$ resource blocks. Note that $a_1+a_2=\alpha$ since $w_{{\text{UL}},i}+w_{{\text{DL}},i}\leq 1, i\in \{1,2\}$. The scheduler activates $\mathbf{v}_{0,1},\mathbf{v}_{1,0},\mathbf{v}_{0,2}$,and $\mathbf{v}_{2,0}$ for $a_3= w_{{\text{DL}},1}-a_2, a_4=w_{{\text{UL}}_1}-a_1, a_5= w_{{\text{DL}},2}-a_1$ and $a_6=w_{{\text{UL}},2}-a_2$ fraction of the resource blocks. It can be verified that this is a valid allocation since $a_i\geq 0$ and $\sum_{i\in [6]}a_i=1$, and all of the temporal demand constraints are satisfied. For $n>2$, a round robin scheduler can be constructed using the same idea as in the previous case. First, the round robin scheduler operates in the FD mode for   $\alpha= \sum_{i\in [n]} {w}_{{\text{UL}},i}+\sum_{i\in [n]} {w}_{{\text{DL}},i}- 1$ fraction of the time-slots. Then, it operates in the HD mode to satisfy the remaining temporal demands. More precisely, the scheduler activates $\mathbf{v}_{1,2}$ for $a_{1,2}=min(\alpha, w_{{\text{UL}},1}, w_{{\text{DL}},2})$ fraction of the time. Then, it activates $\mathbf{v}_{1,3}$ for $a_{1,3}=min(\alpha-a_{1,2}, w_{{\text{UL}},1}-a_{1,2}, w_{{\text{DL}},3})$ fraction of the time. The scheduler proceeds in this manner until $\alpha$ fraction of the resource blocks are scheduled. The fact that the process can be continued until $\alpha$ fraction of the resource blocks is guaranteed from the bounds in the theorem statement. The scheduler proceeds by activating the users in the HD mode until the temporal constraints are satisfied. The proof for $k>0$ follows by similar arguments and is omitted due to space limitations. 

\section{Proof of Theorem \ref{th:2}}
\label{app:th:2}
In the first step, we prove that if $({w}_{\text{DL}}^n,{w}_{\text{UL}}^n)$ satisfies  Equation 
\eqref{eq:feasib}, then it is feasible. Let $(a_{i,j}, a_{0,j}, a_{0,i}), i,j \in [n]$ be the corresponding coefficients satisfying Equation \eqref{eq:feasib}. The round robin strategy characterized by $Q_t(R^{n\times n\times t})= \mathcal{V}_{i_t,j_t}, t\in [s]$, where $(i_t,j_t)$ is the unique index for which the inequality $ \sum_{i\leq i_t,j< j_t} sa_{i,j} +1 \leq t\leq \sum_{i\leq i_t,j\leq j_t} sa_{i,j}$ achieves temporal fairness in the scheduling window-length $s$. Conversely, assume that $({w}_{\text{UL}}^n,{w}_{\text{UL}}^n)$  is feasible, then it is straightforward to see that a round robin strategy satisfying the temporal fairness constraints exists. Let $(a_{i,j}, a_{0,j}, a_{0,i}), i,j \in [n]$ be the shares of the corresponding virtual users for this scheduling strategy. Then, we argue that the coefficients $(a_{i,j}, a_{0,j}, a_{0,i}), i,j \in [n]$ along with temporal demand vectors $({w}_{\text{DL}}^n,{w}_{\text{UL}}^n)$ satisfy Equation \eqref{eq:feasib}. The first two eqaulities hold since the UL and DL temporal demand are satisfied and and the next three equations hold by the definition of the virtual user temporal shares.

\section{Proof of Theorem \ref{th:neq:normal}}
\label{thm-prf:equality}
 First, we prove that if a threshold strategy exists which i) satisfies the temporal constraints, and ii) for which $\lambda_{{\text{UL}},i}, \lambda_{{\text{DL}},i}\in [-2M,2M], \forall i\in [n]$ , then it is optimal, where $M$ is a limited upper-bound on the performance value of any virtual user. Fix $\epsilon>0$. Let $\epsilon'=2nM\epsilon$. 
Let $\widehat{Q}\in \mathcal{Q}_{TBS}$ be a TBS characterized by the threshold vectors $(\lambda_{\text{UL}}^n,\lambda_{\text{DL}}^n)\in [-2M,2M]^n$ and let $Q$ be an arbitrary scheduling strategy. From \eqref{Def:tem_fair} we know that $|A_{i}^{Q}- w_i|\leq \epsilon, \forall i\in [n]$. Also, by assumption, $\lambda_i\leq M, \forall i\in [n]$. As a result, $\lambda_i(A_{i}^{Q}- w_i)+\frac{\epsilon'}{n}\geq 0, \forall i\in [n]$.
We have, \\
\begin{align*}
U^{Q} 
&\leq U^{Q}+\sum_{i=1}^n\big(\lambda_{{\text{UL}},i}(A_{{\text{UL}},i}^{Q}- w_{{\text{UL}},i})\big)+ \sum_{i=1}^n\big(\lambda_{{\text{DL}},i}(A_{{\text{DL}},i}^{Q}- w_{{\text{DL}},i})\big)
+\epsilon' \notag\\
&\leq 
\liminf_{t\rightarrow \infty}
\Bigg[
\frac{1}{t}\sum^t_{k=1}
\sum_{i,j\in[n]\cup \{0\}} 
\Big(R_{i,j,k}
\mathbbm{1}_{\big
\{Q_k(R^{n\times n\times k})=\mathbf{v}_{i,j}\big\}}\Big)\Bigg] 
+\sum_{i=1}^n
\lambda_{{\text{UL}},i}\cdot\liminf_{t\rightarrow \infty}
\frac{1}{t}\Bigg[
\sum^t_{k=1}\sum_{j\in [n]\cup \{0\}}
\Big(
\mathbbm{1}_{\big
\{\mathbf{v}_{i,j}= Q_k({R}^{n\times n\times k})\big\}}\Big)\Bigg]
\\&~~~
+\sum_{i=1}^n
\lambda_{{\text{DL}},i}\cdot\liminf_{t\rightarrow \infty}
\frac{1}{t}\Bigg[
\sum^t_{k=1}\sum_{j\in [n]\cup\{0\}}
\Big(
\mathbbm{1}_{\big
\{\mathbf{v}_{j,i}= Q_k({R}^{n\times n \times k})\big\}}\Big)\Bigg]
-\sum_{i\in [n]\cup \{0\}}(\lambda_{{\text{UL}},i}w_{{\text{UL}},i}+\lambda_{{\text{DL}},i}w_{{\text{DL}},i})+\epsilon'
\\
&\stackrel{(a)}{\leq}
\liminf_{t\rightarrow \infty}
\Bigg[
\frac{1}{t}\sum^t_{k=1}
\sum_{i,j\in[n]\cup \{0\}} 
\Big(R_{i,j,k}
\mathbbm{1}_{\big
\{Q_k(R^{n\times n\times k})=\mathbf{v}_{i,j}\big\}}\Big)
+\sum_{i,j\in [n]\cup \{0\}}
\Big(
(\lambda_{{\text{UL}},i}+\lambda_{{\text{DL}},j})
\mathbbm{1}_{\big
\{\mathbf{v}_{i,j}\in Q_k({R}^{n\times n\times k})\big\}}\Big)\Bigg]
\\&~~~
-\sum_{i\in [n]\cup \{0\}}(\lambda_{{\text{UL}},i}w_{{\text{UL}},i}+\lambda_{{\text{DL}},i}w_{{\text{DL}},i})+\epsilon'
\notag\\
&=\liminf_{t\rightarrow\infty}
\frac{1}{t}\Bigg[
\sum^t_{k=1}
\sum_{i,j\in[n]\cup \{0\}}  \Big(R_{i,j,k}+
\lambda_{{\text{UL}},i}+\lambda_{{\text{DL}},j}\Big)
\mathbbm{1}_{\big
\{\mathbf{v}_{i,j}\in {Q}_k({R}^{n\times n\times k})\big\}}\Bigg]
-\sum_{i\in [n]\cup \{0\}}(\lambda_{{\text{UL}},i}w_{{\text{UL}},i}+\lambda_{{\text{DL}},i}w_{{\text{DL}},i})+\epsilon'\notag\\
&\stackrel{(b)}{\leq}\liminf_{t\rightarrow\infty}
\frac{1}{t}\Bigg[
\sum^t_{k=1}
\sum_{i,j\in[n]\cup \{0\}}  \Big(R_{i,j,k}+
\lambda_{{\text{UL}},i}+\lambda_{{\text{DL}},j}\Big)
\mathbbm{1}_{\big
\{\mathbf{v}_{i,j}\in \widehat{Q}_k({R}^{n\times n\times k})\big\}}\Bigg]
-\sum_{i\in [n]\cup \{0\}}(\lambda_{{\text{UL}},i}w_{{\text{UL}},i}+\lambda_{{\text{DL}},i}w_{{\text{DL}},i})+\epsilon'\notag\\
&\stackrel{(c)}{=}
\liminf_{t\rightarrow \infty}
\Bigg[
\frac{1}{t}\sum^t_{k=1}
\sum_{i,j\in[n]\cup \{0\}} 
\Big(R_{i,j,k}
\mathbbm{1}_{\big
\{\widehat{Q}_k(R^{n\times n\times k})=\mathbf{v}_{i,j}\big\}}\Big)\Bigg] 
+\sum_{i=1}^n
\lambda_{{\text{UL}},i}\cdot\liminf_{t\rightarrow \infty}
\frac{1}{t}\Bigg[
\sum^t_{k=1}\sum_{j\in [n]\cup \{0\}}
\Big(
\mathbbm{1}_{\big
\{\mathbf{v}_{i,j}= \widehat{Q}_k({R}^{n\times n\times k})\big\}}\Big)\Bigg]
\\&~~~+
\sum_{i=1}^n
\lambda_{{\text{DL}},i}\cdot\liminf_{t\rightarrow \infty}
\frac{1}{t}\Bigg[
\sum^t_{k=1}\sum_{j\in [n]\cup\{0\}}
\Big(
\mathbbm{1}_{\big
\{\mathbf{v}_{j,i}= Q_k({R}^{n\times n \times k})\big\}}\Big)\Bigg]
-\sum_{i\in [n]\cup \{0\}}(\lambda_{{\text{UL}},i}w_{{\text{UL}},i}+\lambda_{{\text{DL}},i}w_{{\text{DL}},i})+\epsilon'
\notag\\
&\leq U^{\widehat{Q}}+\underbrace{\sum_{i=1}^n\big(\lambda_{{\text{UL}},i}(A_{{\text{UL}},i}^{\widehat{Q}}- w_{{\text{UL}},i})+
\lambda_{{\text{DL}},i}(A_{{\text{DL}},i}^{\widehat{Q}}- w_{{\text{DL}},i})
\big)}_{\leq \epsilon'}+\epsilon' 
\leq U^{\widehat{Q}}+2\epsilon',
\end{align*}
where (a) holds since limit inferior satisfies supper-additivity, (b) holds due to the rearrangement inequality, and finally, (c) follows from the existence of the limit inferior. As a result:
\begin{align*}
U^{Q} \leq U^{\widehat{Q}}+2\epsilon', \forall \epsilon>0, 
\Rightarrow U^Q\leq U^{\widehat{Q}}.
\end{align*} 
The proof of the existence of the threshold vector follows by similar arguments as in the proof of Theorem 1 in \cite{general-framework-early-access}. We provide an outline of the proof for the case when $\sum_{i\in [n]}w_{UL,i}>1$ and $\sum_{i\in [n]}w_{DL,i}>1$. We use the following Lemma: 
\begin{Lemma}[\textbf{Avoiding Cones Conditions \cite{fonda2016generalizing}}]

Let $n\in \mathbb{N}$. Consider the set of continuous functions $f_i: \mathbb{R}^n \rightarrow \mathbb{R}, i \in [n]$. Assume that for each function $f_i, i\in[n]$, there are positive reals $M^+_i$ and $M^-_i$ such that i) for any point $x^n$ such that  $x_i=M^+_i$, either the function $f_i$ is positive or $\exists j\neq i: f_j(x^n)\neq 0$, and ii) For any point $x^n$ such that $x_i=-M^-_i$, either the function $f_i$ is negative or $\exists j\neq i: f_j(x^n)\neq 0$. Then, the function $f^n=(f_1,f_2,\cdots,f_n)$ has a root in the 
 n-dimensional cube $\prod_{i=1}^n[-M^-_i,M^+_i]$. Alternatively:
\begin{align*}
 \exists x^*_1,\ldots,x^*_n\in  \prod_{i=1}^n[-M^-_i,M^+_i] : f_i(x^*_1,\ldots,x^*_n)=0,\forall i\in[n]. 
\end{align*}
\label{Lem:AE}
\end{Lemma}

Take $f_{UL,i}(\lambda_{UL}^n,\lambda_{DL}^n)\triangleq A_{UL,j}^{Q_{TBS}}-w_{UL,j}, \forall i\in [n]$ and $f_{DL,i}(\lambda_{UL}^n,\lambda_{DL}^n)\triangleq A_{DL,j}^{Q_{TBS}}-w_{DL,j}, \forall i\in [n]$. First we construct $M_{UL,i}^-$ and $M_{DL,i}^-, i\in [n]$ satisfying the conditions in Lemma \ref{Lem:AE}. 
Note that by assumption $e_{UL}\triangleq  \frac{\sum_{i\in [n]}w_{UL,i}-1}{n}>0$ and $e_{DL}\triangleq \frac{\sum_{i\in [n]}w_{DL,i}-1}{n}>0$. Furthermore, it is straightforward to show that there exists $\alpha_{UL}^n, \alpha_{DL}^n>0$ such that $\sum_{i\in [n]}\alpha_{UL,i}=\sum_{i\in [n]}\alpha_{DL,i}=1$ and $w_{UL,i}-\alpha_{UL,i}e_{UL},w_{DL,i}-\alpha_{DL,i}e_{DL}>0, i\in [n]$. Define $w'_{UL,i}=w_{UL,i}-\alpha_{UL,i} e_{UL}$, $w'_{DL,i}=w_{DL,i}-\alpha_{DL,i} e_{DL},i\in [n]$. Then, by construction,  $\sum_{i\in [n]} w'_{UL,i}=\sum_{i\in [n]} w'_{DL,i}=1$. So, the temporal demands $({w'}^n_{UL},{w'}^n_{DL})$ can be satisfied by activating a single user at each slot. 
Note that if $\lambda_{UL,i},\lambda_{DL,i} \leq -M, \forall i\in [n]$, then only the individual users will be chosen by the threshold strategy. The reason is that in this case the scheduling measures for the HD virtual users are larger than that of joint users 
with probability one. Let $\lambda^-_{UL,i}, \lambda^-_{DL,i}\in [-2M,-M], i\in [n]$ be the thresholds for an HD scheduler (with virtual users $\mathcal{V}_{0,i}, \mathcal{V}_{i,0}, i\in [n]$)  satisfying the temporal constraints $({w'}^n_{UL},{w'}^n_{DL})$.  
Then, $M_{UL,i}^-=\lambda_{UL,i}, M_{DL,i}^-=\lambda_{DL,i}, i\in [n]$ satisfy the condition that $f_{UL,i}(\lambda_{UL}^n,\lambda_{DL}^n)<0,  \lambda_{UL,i}=M_{UL,i}^-$, and $f_{DL,i}(\lambda_{UL}^n,\lambda_{DL}^n)<0,  \lambda_{DL,i}=M_i^-$ $\forall i\in [n]$. Furthermore, $M^+_{UL,i}=M^+_{DL,i}=2M, i\in [n]$ satisfy the conditions in Lemma \ref{Lem:AE}. The reason is that for instance if $M^+_{UL,i}=2M$, then the scheduler only activates FD virtual users or HD users in UL. Then, at each time-slot, either the $i$th user is chosen in UL or there exists user $u_j$ for which  $f_{UL,j}(\lambda_{UL}^n,\lambda_{DL}^n)>0$ since $\sum_{i\in [n]}w_{UL,i}<2$. So, we have shown that the conditions for Lemma \ref{Lem:AE} are satisfied which proves the existence of threshold values for which $f_{UL,i}(\lambda_{UL}^n,\lambda_{DL}^n)= A_{UL,j}^{Q_{TBS}}-w_{UL,j}=0, \forall i\in [n]$ and $f_{DL,i}(\lambda_{UL}^n,\lambda_{DL}^n)= A_{DL,j}^{Q_{TBS}}-w_{DL,j}=0, \forall i\in [n]$.

\end{appendices}
\bibliographystyle{IEEEtran}
\bibliography{reference}

\end{document}